\documentstyle[prl,aps,amssymb,epsfig,graphicx,tabularx]{revtex}

\def\be{\begin{equation}}
\def\ee{\end{equation}}
\def\bea{\begin{eqnarray}}
\def\eea{\end{eqnarray}}

\newcommand{\ket}[1]{|\kern.3ex#1\kern.3ex\rangle}
\newcommand{\bra}[1]{\langle\kern.3ex #1 \kern.3ex|}

\newcommand{\mean}[1]{\left\langle #1 \right\rangle} 
\newcommand{\smean}[1]{\langle #1 \rangle} 

\newcommand{\tr}[1]{\mathop{\mathrm{Tr}}\nolimits\left\{ #1 \right\}}  
\newcommand{\str}[1]{\mathop{\mathrm{Tr}}\nolimits\{ #1 \}}  
  
\def\D{{\rm d}}                  

\def\vep{\varepsilon}

\begin{document}

\twocolumn[\hsize\textwidth\columnwidth\hsize\csname@twocolumnfalse\endcsname

\title{Effect of incoherent scattering on shot noise correlations 
       in the quantum Hall regime}

\author{Christophe Texier and Markus B\"uttiker}

\address{D\'epartement de Physique Th\'eorique. Universit\'e de Gen\`eve. \\
         24, quai Ernest Ansermet. CH-1211 Gen\`eve 4. Switzerland.}

\date{March 17, 2000}

\maketitle

\begin{abstract}
We investigate the effect of incoherent scattering in a 
Hanbury Brown and Twiss situation with electrons in edge states of a 
three-terminal conductor submitted to a strong perpendicular magnetic field. 
The modelization of incoherent scattering is performed by introducing an 
additional voltage probe through which the current is kept equal to zero 
which causes voltage fluctuations at this probe. It is shown that inelastic
scattering can lead in this framework to positive correlations, whereas 
correlations remain always negative for quasi-elastic scattering.
\end{abstract}

\twocolumn
\vskip.5pc]
\narrowtext


\section{Introduction}

\begin{figure}
\includegraphics[scale=0.95]{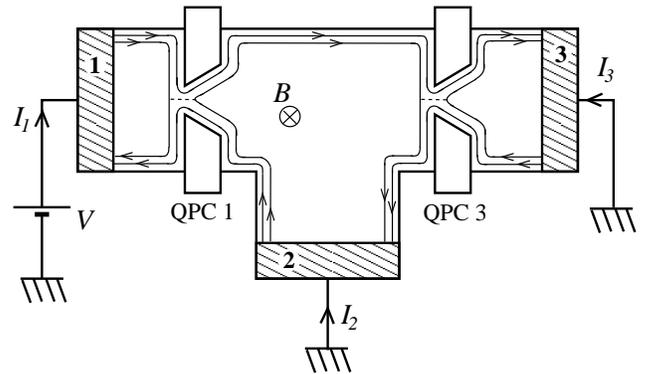}
\caption{Experimental arrangement of \protect\cite{ObeHenStrSchHeiEnsHol00}.
The two edge states are populated at contact 1. One of those channels
is perfectly transmitted the other being partially transmitted at QPC 1.
The transmission may be tuned by an applied gate voltage. QPC 3 splits the 
partially degenerate beam and distributes the current to contacts 2 and 3 
where it is measured. Without inter-edge state scattering, the perfectly
transmitted channel does not manifest itself on the correlations.
\label{fig:exp}}
\end{figure}

The effect of incoherent scattering on transport and noise in 
mesoscopic structures has been of interest in a number of 
previous works (for review see \cite{JonBee96,BlaBut00}). 
In this paper we are interested in the effect of inelastic and quasi-elastic
scattering on the correlations of the current in a structure submitted to a
strong magnetic field so that the current is carried by edge states propagating
along the boundary of the sample (see figure \ref{fig:exp}).

To demonstrate the reality of edge states 
\cite{Hal82} (despite their small contribution
to the overall density of states) the possibility of creating a 
non-equilibrium population \cite{But88} has been crucial.
Several experiments have investigated  the equilibration of edge states
selectively populated with the help of various contacts by transport 
measurements
\cite{WeeWilKouHarWilFoxHar89,KomHirSasFuj90,AlpMcEWheSac90,MulWeiKhavKlKocNicSchLos92}. 
A natural problem is to investigate the effect of inter-edge state scattering 
on the noise on such structures. 

Recently, Hanbury Brown and Twiss (HBT) experiments studying current
correlations using partially degenerate stream of 
fermions were performed \cite{HenObeStrHeiEnsHolSch99,ObeHenStrSchHeiEnsHol00},
starting from the idea \cite{But90a} of using the edge states in a
conductor submitted to a strong magnetic field. 
A Y-structure is discussed in Ref. \cite{MarLan92} and a HBT experiment 
without a magnetic field has also been realized \cite{OliKimLiuYam99}.
Here we are interested in the experimental arrangement of Oberholzer 
{\it et al.} \cite{ObeHenStrSchHeiEnsHol00} which is depicted in figure 
\ref{fig:exp}, with the 
only difference that the magnetic field, in this experiment, was adjusted in 
such a way that only one spin-degenerate edge state (filling factor $\nu=2$) 
carries the current, and not two, as it is represented on the figure.
Contact 1, which is at a potential $eV$ above the potentials of the two other
contacts, acts as a carrier source and contacts 2 and 3 as 
detectors for the beams of electrons. The incident beam is splitted at two 
quantum point contacts (QPC) which can be tuned by applied voltages. In the 
following we will denote by $T_1$ and $R_1=1-T_1$ the transmission and 
reflection coefficients at QPC at contact 1, and introduce $T_3$ and $R_3$ 
for QPC at contact 3.
The quantity of interest in this experiment, revealing statistical properties
of the carriers, is the correlation $S_{23}=\mean{\Delta I_2\Delta I_3}$ 
between the current at the two contacts.
Compared to the initial experiment by Henny {\it et al.} 
\cite{HenObeStrHeiEnsHolSch99} the experiment by Oberholzer {\it et al.}
\cite{ObeHenStrSchHeiEnsHol00} introduces fluctuations in the incident beam 
and thus the current correlations are not determined already 
from current conservation alone.
In the following we will be interested in the case where the distance between
the QPC's is long so that incoherent processes may occur along this edge.
If only one edge state is populated, phase breaking processes are of little 
interest since $S_{23}$ measures quantum-statistical properties of the carriers
revealed by the separation of the beam at QPC 3, and what happens before 
this separation is of little importance. 
We can indeed check that the introduction of incoherent processes within
the framework that will be used in the following does not affect the 
auto-correlations (noise) and the correlations if only one edge state is 
present.

In contrast, when several edge states are populated, inter-edge state 
scattering along one boundary may cause a redistribution of the carriers 
between the edge states and thus modify the 
noise properties of each channel. This is the situation that will interest
us in this paper. 
Inter-edge state scattering centers were investigated 
experimentally very recently \cite{WooValMcEKadMarGos00}.
We will not take into account the spin degeneracy which only
would multiply the conductances and the noise by a factor 2.

The inelastic \cite{But86,BeeBut92,JonBee96} and quasi-elastic scattering 
\cite{JonBee96,LanBut97} are modeled by introducing an additional probe at the 
edge along which incoherent scattering occurs (see figure \ref{fig:dephase}). 
We have to impose that the current through this probe is zero at any
time. This approach, followed in a number of works 
\cite{JonBee96,But86,LanBut97,DamPas90,And96,GagMas96,MorJauFle99,LiuChu00} 
(see \cite{BlaBut00} for a review), has the advantage to reduce the problem 
to the study of coherent scattering in a conductor with one additional contact.
The method will be recalled in more detail in the following.

In the first section we are interested in the situation where only coherent 
scattering is present. In the following section we describe the influence 
of incoherent scattering along the long edge of the sample.


\section{Coherent Scattering}

We first discuss the situation when only elastic scattering is present
in the system. We consider the situation depicted in figure \ref{fig:exp} 
where the magnetic field is chosen so that two edge states are populated.
The edge state associated with the lowest Landau level (LL) is perfectly 
transmitted at the two QPC's whereas the second edge state associated with the 
second LL is only partially transmitted at the QPC's.
We are interested in the current correlations between contacts 2 and 3.

The noise spectrum is defined as 
$S_{\alpha\beta}(\omega)2\pi\delta(\omega+\omega')=
\smean{\Delta\hat I_\alpha(\omega)\Delta\hat I_\beta(\omega')
+\Delta\hat I_\beta(\omega')\Delta\hat I_\alpha(\omega)}$
with $\Delta\hat I_\alpha(\omega)=
\hat I_\alpha(\omega)-\smean{\hat I_\alpha(\omega)}$, where 
$\hat I_\alpha(\omega)$ is the Fourier transform of the current 
operator at contact $\alpha$. 
(For a recent presentation of the formalism and notations, see 
\cite{BlaBut00}).
The zero frequency limit will be denoted: 
$S_{\alpha\beta}=S_{\alpha\beta}(\omega=0)$.
Following the scattering approach, we construct the scattering 
matrix for the system to calculate the noise \cite{But92,BlaBut00}:
\be\label{noise}
S_{\alpha\beta}=\frac{2e^2}{h}\int\D E\, \sum_{\gamma,\lambda}
\tr{A_{\gamma\lambda}^\alpha A_{\lambda\gamma}^\beta} f_\gamma(1-f_\lambda)
\:,\ee
where the matrices $A$'s are related to the on-shell $S$-matrix: 
$A_{\gamma\lambda}^\alpha=\delta_{\alpha\gamma}\delta_{\alpha\lambda}
-s^\dagger_{\alpha\gamma}(E)s_{\alpha\lambda}(E)$, $f_\alpha$ being the 
Fermi-Dirac distribution at the contact $\alpha$. We consider the 
case of zero temperature and will discuss at the end the effect of temperature
on some of the results.

Using this expression and the $S$-matrix we can compute the correlations 
between the currents at contacts $2$ and $3$, which is eventually found to 
be \cite{ObeHenStrSchHeiEnsHol00}:
\be\label{ObRes}
S_{23}=-\frac{2e^2}{h} |eV| R_3 T_3 T_1^2
\:,\ee
where $R_1$, $T_1$ and $R_3$, $T_3$ are the reflection and transmission 
coefficients for the edge state corresponding to the second LL at QPC 1 and 3,
respectively. Since the first edge state is totally transmitted through QPC's, 
it carries a noiseless current and does not contribute to the noise or to 
the correlations.
The $R_3 T_3$ factor is the usual partition factor due to the separation of 
the electron beam at QPC 3. The $T_1^2$ term is due to the fact that the
processes leading to correlations between contacts 2 and 3 involve two 
electrons transmitted through the QPC 1, which happens with probability 
$T_1^2$.

Before discussing the effect of inelastic and quasi-elastic scattering, we
may consider first the effect of coherent scattering between the two edge 
states along the upper edge. This question might seem academic but it will
be important to have this result in mind, to appreciate the difference in
the correlations $S_{23}$ when the two edge states exchange carriers
coherently or incoherently. 
Let us recall that in the presence of coherent scattering, it was proven 
that correlations between two contacts are always negative as a consequence 
of the fermionic nature of the carriers\footnote{
The proof applies only to conductors which are part of a zero-impedance 
external circuit.} \cite{But92}.
To describe coherent scattering between edge states we introduce the 
probability $\vep$ that an electron, starting in one of the two channels, 
is scattered into the other edge state when it travels between the 
two QPC's. Here we do not need to enter into more details about the
scattering however let us mention that it has been studied within a 
microscopic approach in \cite{OhtOno89,MarFen90}.  
After having constructed the $S$-matrix, 
formula (\ref{noise}) gives the correlations:
\bea
S^{\rm coh}_{23}=-\frac{2e^2}{h} |eV| R_3 \Big[ 
(1-\vep)^2T_3 T_1^2 + \vep^2 T_3 \nonumber\\
+ \vep(1-\vep) (1-2R_3T_1+T_1^2)\Big]
\:.\eea
Let us now discuss a few limiting cases. If $\vep=0$ (no inter-edge state
elastic scattering) we fall back to the 
previous situation (\ref{ObRes}).  If $\vep=1$, the two 
edge states are perfectly exchanged between the QPC's and the current 
separated by QPC 3 issues from the first edge state, perfectly transmitted
at QPC 1 and we have simply $S^{\rm coh}_{23}=-\frac{2e^2}{h} |eV| R_3 T_3$. 
Note that the outer edge state in contact 3 is not noiseless but these 
fluctuations are not correlated with fluctuations at contact 2 and then do 
not contribute to $S_{23}$ since this edge state is not splitted at QPC 3.

If $T_1=0$ the correlation is 
$S^{\rm coh}_{23}=-\frac{2e^2}{h} |eV|[R_3T_3\vep^2 + \vep(1-\vep)R_3]$. 
The first term is the partition noise
for the beam separated at QPC 3; this term is proportional to the probability
$\vep^2$ that two electrons are transmitted from the first edge state to the 
second. The second term is the partition noise due to the separation of the 
beam between QPC's. In this process, after the separation of the beam, one 
electron is transmitted with probability $1$ in the first edge state towards 
contact 3 and the other is reflected with probability $R_3$ towards contact 2.

Another interesting case occurs for $T_3=0$. Then the two edge states
are perfectly separated at QPC 3. The first edge state flows towards contact
3 whereas the second goes to contact 2. This situation is 
particularly interesting since it allows us to study separately the noise 
spectrum of each edge state after having travelled along the upper edge of 
the conductor. In the absence
of scattering between edge states, the currents remain uncorrelated, whereas
if redistribution of charges occurs between edge states, it may induce
correlations. For coherent scattering, we find: 
$S^{\rm coh}_{23}=-\frac{2e^2}{h}|eV|\vep(1-\vep)R_1^2$. 
The only separation of the beam that can cause
correlations occurs along the edge with usual partition factor $\vep(1-\vep)$; 
the factor $R_1^2$ ensures that, if the currents in the two channels are 
noiseless before exchanging carriers along the upper edge, they remain 
noiseless and uncorrelated.
Let us also give the currents auto-correlations:
$S^{\rm coh}_{22}=\frac{2e^2}{h} |eV| [R_1T_1(1-\vep)^2+\vep(1-\vep)R_1]$
and
$S^{\rm coh}_{33}=\frac{2e^2}{h} |eV| [R_1T_1\vep^2+\vep(1-\vep)R_1]$.
The noise spectral densities have contributions from the splitting of the 
beam at the first QPC and from the redistribution of the charges between 
the QPC's. 
Let us finally mention the value of the noise when $\vep=1/2$, {\it i.e.} 
when the average currents in the two edge states are equally equilibrated 
due to scattering, then
\be\label{S23coh}
S^{\rm coh}_{23}=-\frac{e^2}{h} |eV| \frac{R_1^2}{2}
\hspace{0.5cm} \mbox{ for } T_3=0,\ \vep=1/2
\ee
and
\be
S^{\rm coh}_{22}=S^{\rm coh}_{33}=\frac{e^2}{h} |eV| \frac{1+T_1}{2}
{R_1}
\:.\ee

Next we proceed to treat incoherent scattering.


\section{Introduction of incoherent scattering}

\begin{figure}
\begin{center}
\includegraphics[scale=1]{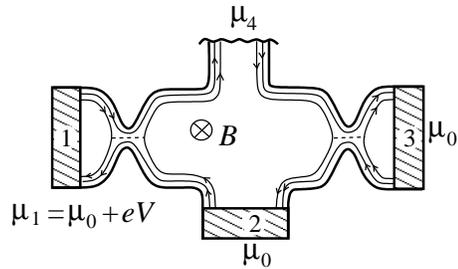}
\end{center}
\caption{Modelization of phase breaking processes along the upper edge.
         \label{fig:dephase}}
\end{figure}

As we have already evoked, the presence of inelastic or 
quasi-elastic scattering will be treated by adding a fictitious contact
at the edge along which incoherent scattering is expected to occur
\cite{But86,JonBee96,BlaBut00}. Let us remark that the only place where
it can have some influence on the correlations is the edge between the
two QPC's where we have introduced it (see figure \ref{fig:dephase}).
Since the current is obviously conserved along the edge, we have to impose 
that the current $I_4$ through this contact is zero at any time. The 
consequence is that the potential $\mu_4$ at this probe is fluctuating.

The average currents are related to the potentials by the conductance:
$\mean{I_\alpha}=\frac1e\int\D E\,\sum_\beta G_{\alpha\beta}(E)
\bar f_\beta(E)$, where we recall that
$G_{\alpha\beta}=\frac{e^2}{h}(
N_\alpha\delta_{\alpha\beta}-\str{s^\dagger_{\alpha\beta}s_{\alpha\beta}})$,
$N_\alpha$ being the number of open channels at contact $\alpha$.
We first discuss the inelastic case \cite{But86} for which we require that 
the current is zero on average $\mean{I_4}=0$ and that the average 
distribution function at the fictitious contact is an equilibrium Fermi 
distribution. This leads to the following expression for the average 
chemical potential at contact $4$:
\be\label{ap}
\bar\mu_4=\mu_0 + \frac{1+T_1}{2} eV
\:.\ee
The fact that $I_4$ is zero on average determines the average 
distribution function at contact $4$. We have also to ensure that the 
fluctuating part of $I_4$ remains zero.
At contact $\alpha$ the current may be written as 
$I_\alpha
=\frac1e\int\D E\,\sum_\beta G_{\alpha\beta}f_\beta + \delta I_\alpha$, where
$\delta I_\alpha$ is the intrinsic part of the fluctuations, whose spectrum
is given by (\ref{noise}). We now write the currents as 
$I_\alpha=\mean{I_\alpha}+\Delta I_\alpha$. Imposing that the fluctuating 
current is zero $\Delta I_4=0$ leads to the relation
$\delta I_4=-\frac1eG_{44}(\mu_4-\bar\mu_4)$ which gives the
expression of the fluctuating part of the current:
\be\label{Di}
\Delta I_\alpha=\delta I_\alpha - \frac{G_{\alpha4}}{G_{44}}\delta I_4
\:,\ee
the first term corresponds to intrinsic fluctuations of the current and 
the second term to the fluctuations due to the existence of a fluctuating 
potential at contact $4$. 

In the quasi-elastic case \cite{JonBee96}
we impose that not only the total current
$I_4=\int\D E\,j_4(E)$ is zero on average but the contribution to the current
of the states of energy $E$ is zero $\mean{j_4(E)}=0$, which gives the 
averaged distribution function at contact $4$:
\be\label{ad}
\bar f_4(E) = \frac{1+T_1}{2} f_1(E) + \frac{R_1}{2} f_2(E)
\:.\ee

\begin{figure}
\begin{center}
\includegraphics[scale=0.95]{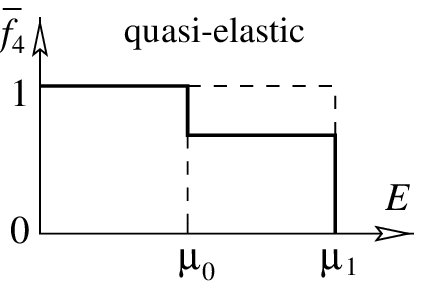}
\hspace{0.25cm}
\includegraphics[scale=0.95]{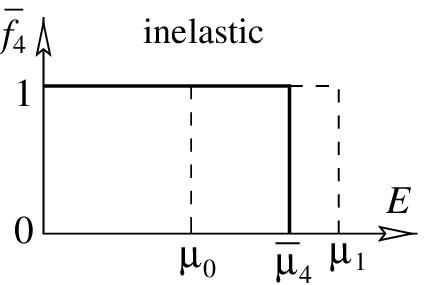}
\caption{Distribution functions at the fictitious contact.\label{fig:dist}}
\end{center}
\end{figure}

The distribution function at the fictitious probe is plotted in the two 
cases in figure \ref{fig:dist}. Since the integral of these two distribution
functions are equal, the currents in the two edge states are equally 
equilibrated by these two kinds of incoherent scattering. Indeed, if we 
compute the average currents for $T_3=0$, when the two edge states are 
directed each to a different contact, we find in the two cases:
$\mean{I_2}=\mean{I_3}=-\frac{e^2}{h}\frac{1+T_1}{2}V$.

In the quasi-elastic case, the distribution function $f_4$ is itself a 
fluctuating quantity. Imposing that the contribution of the states of energy 
$E$ to the fluctuating part of the current at the fictitious contact is 
zero, $\Delta j_4(E)=0$, leads to a 
relation between the fluctuating part of the distribution $f_4-\bar f_4$ 
and the contribution of those states to the intrinsic noise $\delta j_4(E)$. 
This relation is of the same form as (\ref{Di}):
$\Delta j_\alpha(E)=\delta j_\alpha(E)
-\frac{G_{\alpha4}}{G_{44}}\delta j_4(E)$.

Finally we find for the two different kinds of incoherent scattering the 
current correlations
$S^{\rm in,qe}_{\alpha\beta}=\mean{\Delta I_\alpha\Delta I_\beta}$:
\be
S^{\rm in,qe}_{\alpha\beta} = S_{\alpha\beta} - \frac{G_{\alpha4}}{G_{44}}
S_{\beta4} - \frac{G_{\beta4}}{G_{44}}S_{\alpha4} + 
\frac{G_{\alpha4}G_{\beta4}}{G_{44}^2}S_{44}
\:.\ee
They involve the conductances $G_{\alpha\beta}$ and the intrinsic 
correlations $S_{\alpha\beta}=\mean{\delta I_\alpha\delta I_\beta}$ 
for the four-terminal conductor of figure \ref{fig:dephase}, calculated with
formula (\ref{noise}) where the distribution function at contact 4 is either 
a step like Fermi function for the potential $\bar\mu_4$ in the inelastic 
case (\ref{ap}), or the average distribution $\bar f_4$ given by (\ref{ad}) 
in the quasi-elastic case (see figure \ref{fig:dist}).

Let us now come to the results. First, for the quasi-elastic case, we find:
\be\label{Sqe}
S^{\rm qe}_{23}=-\frac{e^2}{h}|eV| \frac{R_3}{4}\Big[ 2 T_3(1+T_1)^2 
+ (1+T_3)R_1^2 \Big]
\:.\ee
In particular, in the interesting limit $T_3=0$ where the two edge states are
perfectly separated at QPC 3, we get:
$S^{\rm qe}_{23}=-\frac{e^2}{h}|eV| \frac{R_1^2}{4}$, {\it i.e.} half of the
result (\ref{S23coh}) obtained when scattering between edge states is coherent.
The auto-correlations take the values: 
$S^{\rm qe}_{22}=S^{\rm qe}_{33}=\frac{e^2}{h}|eV|\frac{R_1}{2}
\frac{1+3T_1}{2}$.

\begin{figure}
\begin{center}
\includegraphics[scale=1]{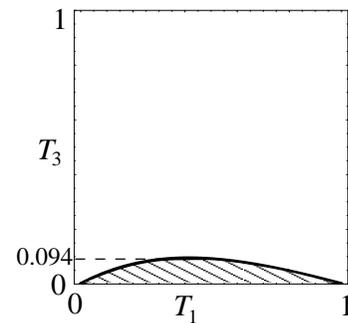}
\caption{Sign of correlations $S^{\rm in}_{23}$ in the presence of inelastic 
scattering. The dashed area corresponds to the region in the parameter space
where $S^{\rm in}_{23}>0$.\label{fig:s23}}
\end{center}
\end{figure}

More surprising is the result obtained for inelastic scattering:
\be\label{Sine}
S^{\rm in}_{23}=-\frac{e^2}{h}|eV| \frac{R_3}{2}
\Big[ 2 T_3(1+T_1) - (1+T_3)R_1T_1 
\Big]
\ee
leading to the possibility of positive correlations, as figure \ref{fig:s23} 
shows (let us recall again that correlations in the presence of coherent 
scattering only are always negative due to the fermionic nature of carriers 
\cite{But92}). If $T_3=0$, we get 
$S^{\rm in}_{23}=+\frac{e^2}{h}|eV|\frac{R_1T_1}{2}$. This result  means 
that this modelization of inelastic scattering gives the possibility for a 
fluctuation of the potential $\mu_4$ to inject in a correlated way in the 
two channels some current.
Let us also mention the result for the noise in this limit:
$S^{\rm in}_{22}=S^{\rm in}_{33}=\frac{e^2}{h}|eV|\frac{R_1T_1}{2}$.

If $T_1=T_3=0$, only one edge state is transmitted at QPC $1$, the currents 
in the two channels are equilibrated by inelastic processes between QPC's, 
and finally the two edge states are separated by QPC 3.
We remark that in this case the correlations and the noise vanish. This shows
that this modelization of inelastic process does not introduce noise when it 
distributes the current of a noiseless channel into two channels.
(This is not the case if coherent or quasi-elastic scattering occurs between 
edge states: $S^{\rm coh}_{23}=-\frac{e^2}{h}|eV|\frac{1}{2}$ and
$S^{\rm qe}_{23}=-\frac{e^2}{h}|eV|\frac{1}{4}$).

Note that if more than two edge states carry the current, still with only one 
partially transmitted and all others being perfectly transmitted, the region in
the paramater space where the correlations are positive diminishes.


Finally we would like to discuss the effect of a finite temperature on the 
result (\ref{Sine}) for $T_3=0$. We will show that for a high enough 
temperature $S^{\rm in}_{23}$ is negative. For a finite temperature $T$, the 
Fermi distributions at the three contacts are smoothed as well as the 
average distribution $\bar f_4$ at the fictitious contact. For $T_3=0$, 
equation (\ref{noise}) gives:
\bea
S^{\rm in}_{23}=\frac{e^2}{h}\bigg[ -k_{B}T
\left(2+{R_1}+{R_1T_1}\right) \hspace{2cm}\nonumber \\
+\frac{R_1T_1}{2}{eV}\coth\frac{eV}{2k_{B}T}\bigg]
\:.\eea
If $T=0$ we indeed recover the positive result 
$S^{\rm in}_{23}=+\frac{e^2}{h}|eV|\frac{R_1T_1}{2}$ for the shot noise, and
if $V=0$ we find 
$S^{\rm in}_{23}=-\frac{2e^2}{h}k_{B}T\left(1+\frac{R_1}{2}\right)$, which is 
the result of the fluctuation-dissipation theorem:
$S^{\rm in}_{23}=2k_{B}T(G^{\rm in}_{23}+G^{\rm in}_{32})$ where
$G^{\rm in}_{\alpha\beta}=G_{\alpha\beta}-\frac{G_{\alpha4}G_{4\beta}}{G_{44}}$
is the conductance of the three-terminal conductor in the presence of 
incoherent scattering.

Let us define $T_c$, the critical temperature above which the correlations
$S^{\rm in}_{23}$ are negative. For small transmission $T_1\ll1$ we find:
$k_{B}T_c\simeq |eV|\frac{T_1}{6}$, and for large transmission $R_1\ll1$:
$k_{B}T_c\simeq |eV|\frac{R_1}{4}$. The transmission that maximizes the 
critical temperature is $T_1=3-\sqrt6\simeq0.55$. In this case we have:
$k_{B}T_c^{\rm max}\simeq
|eV|\frac{5\sqrt6-12}{2(6\sqrt6-12)}\simeq\frac{|eV|}{21.8}$.


\section{Summary}

To summarize we have shown that incoherent scattering can have a strong
effect on the correlations in a HBT situation. In our modelization, inelastic
scattering is responsible for positive correlations in a certain parameter
range. We emphasize that the investigation of the correlations with
the conductor of figure \ref{fig:exp} for $T_3=0$ provides direct 
information about inter-edge state scattering: in the absence of inter-edge
state scattering the correlation $S_{23}$ vanishes, for elastic scattering 
and for quasi-elastic scattering it is negative and for inelastic scattering 
it is positive. In principle it is possible to discriminate also between 
elastic inter-edge state scattering and quasi-elastic scattering: in the 
case of elastic scattering a single parameter ($\vep$) determines the noise
spectra and the conductances if $T_1$ and $T_3$ are known.

We have constructed here only the limiting case of a fictitious contact. 
Partially transmitting contacts would allow to interpolate between fully 
coherent inter-edge state scattering and fully quasi-elastic or fully
inelastic inter-edge state scattering.


\section*{Acknowlegments}

Interest in the topic of this work was stimulated by a question of Bart 
J.~van~Wees.
We acknowledge Yaroslav M.~Blanter and Andrew M.~Martin for very interesting 
discussions. This work was supported by the Swiss National Science Foundation 
and by the TMR Network Dynamics of Nanostructures.



\end{document}